\documentclass[epj]{webofc}
\usepackage[varg]{txfonts} 

\usepackage{xcolor}
\usepackage{cancel}

\wocname{EPJ Web of Conferences}
\woctitle{Strangeness in Quark Matter 2021}
\begin{document}
\selectlanguage{english}
\title{Triple nuclear collisions – a new method to explore the matter properties under new extreme conditions}
%
%
\author{O. V. Vitiuk\inst{1}
        \and
        V. M.~Pugatch\inst{1}
        \and
        K.A. Bugaev\inst{2,3} 
        \and
        P. P. ~Panasiuk\inst{3}
        \and
        N. S. Yakovenko\inst{3}
        \and
        B. E. Grinyuk\inst{2}
        \and
        E. S. Zherebtsova\inst{4,5}
        \and
        M.~Bleicher\inst{6}
        \and
        L.V. ~Bravina\inst{7}
        \and
        A.V. ~Taranenko\inst{4,5}
        \and
        E.E.~Zabrodin\inst{7,8}
}

\institute{Institute for Nuclear Research, NAS of Ukraine, Prospekt Nauki av. 47, 03680 Kyiv, Ukraine
        \and
        Bogolyubov Institute for Theoretical Physics, NAS of Ukraine, Metrologichna str. 14-B, 03680 Kyiv, Ukraine
        \and
        Department of Physics, Taras Shevchenko National University of Kyiv, 03022 Kyiv, Ukraine
        \and
        National Research Nuclear University (MEPhI), Kashirskoe Shosse 31, 115409 Moscow, Russia
        \and
        Institute for Nuclear Research, Russian Academy of Science, 108840 Moscow, Russia
        \and
        ITP, Goethe University, Max-von-Laue-Str. 1, 60438 Frankfurt am Main, Germany
        \and
        University of Oslo, POB 1048 Blindern, N-0316 Oslo, Norway
        \and
        Skobeltsyn Institute of Nuclear Physics, Moscow State University, 119899 Moscow, Russia
}

\abstract{
We suggest to explore an entirely new method to experimentally and theoretically study the phase diagram of strongly interacting matter based on the triple nuclear collisions (TNC). We simulated the TNC using the UrQMD 3.4 model at the beam center-of-mass collision energies $\sqrt{s_{NN}} = 200$ GeV and $\sqrt{s_{NN}} = 2.76$ TeV. It is found that in the most central and simultaneous TNC the initial baryonic charge density is about 3 times higher than the one achieved in the usual binary nuclear collisions at the same energies. As a consequence, the production of protons and $\Lambda$-hyperons is increased by a factor of 2 and 1.5, respectively. Using the MIT Bag model equation we study the evolution of the central cell in TNC and demonstrate that for the top RHIC energy of collision the baryonic chemical potential is 2-2.5 times larger than the one achieved in the binary nuclear collision at the same type of reaction. Based on these estimates, we show that TNC offers an entirely new possibility to study the QCD phase diagram at very high baryonic charge densities.
}

\maketitle

\section{Introduction}
\label{intro_Bugaev}

\begin{figure}[t]
        \centerline{
                \includegraphics[width=70mm]{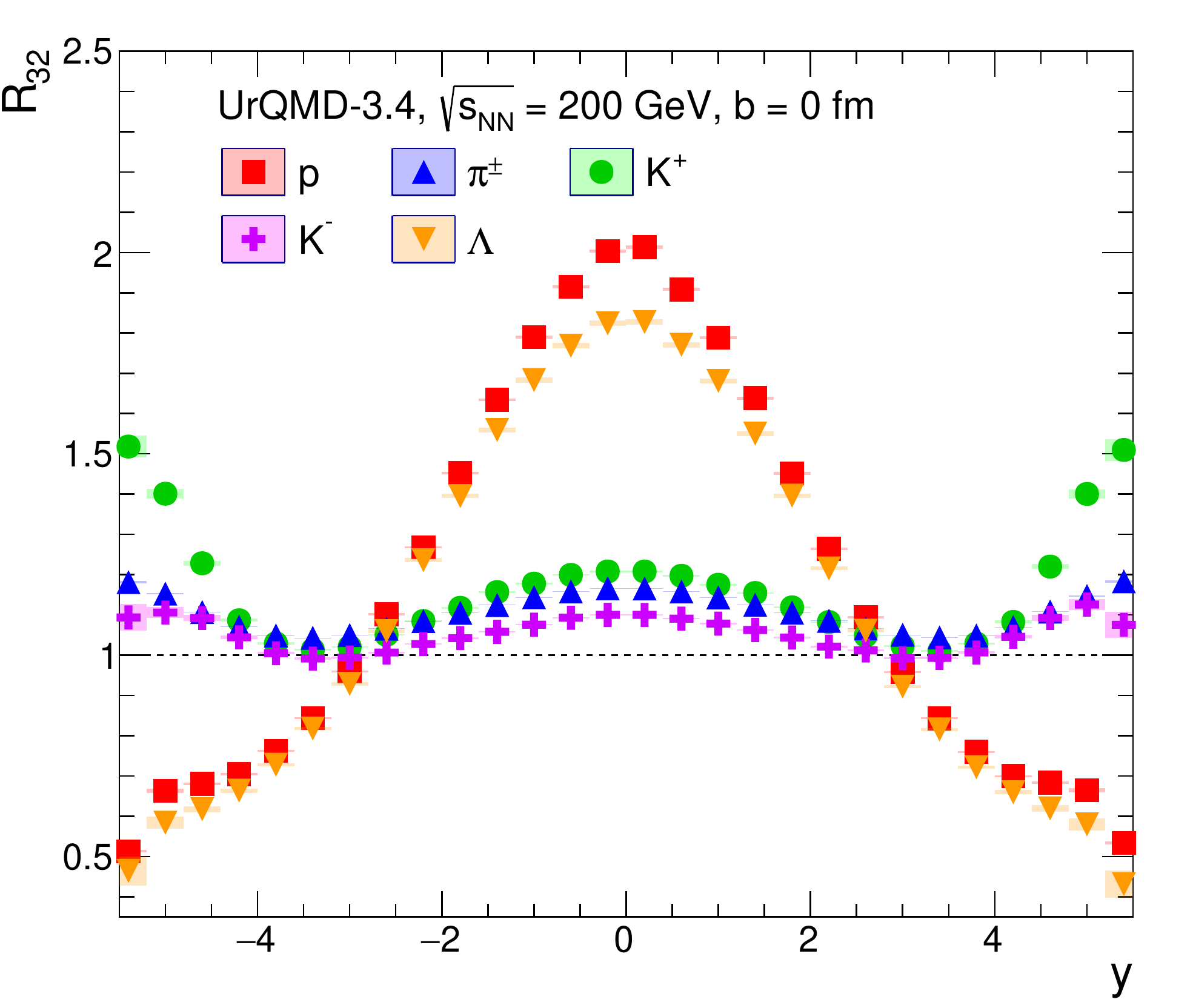} \hspace*{-1.1mm}
                \includegraphics[width=70mm]{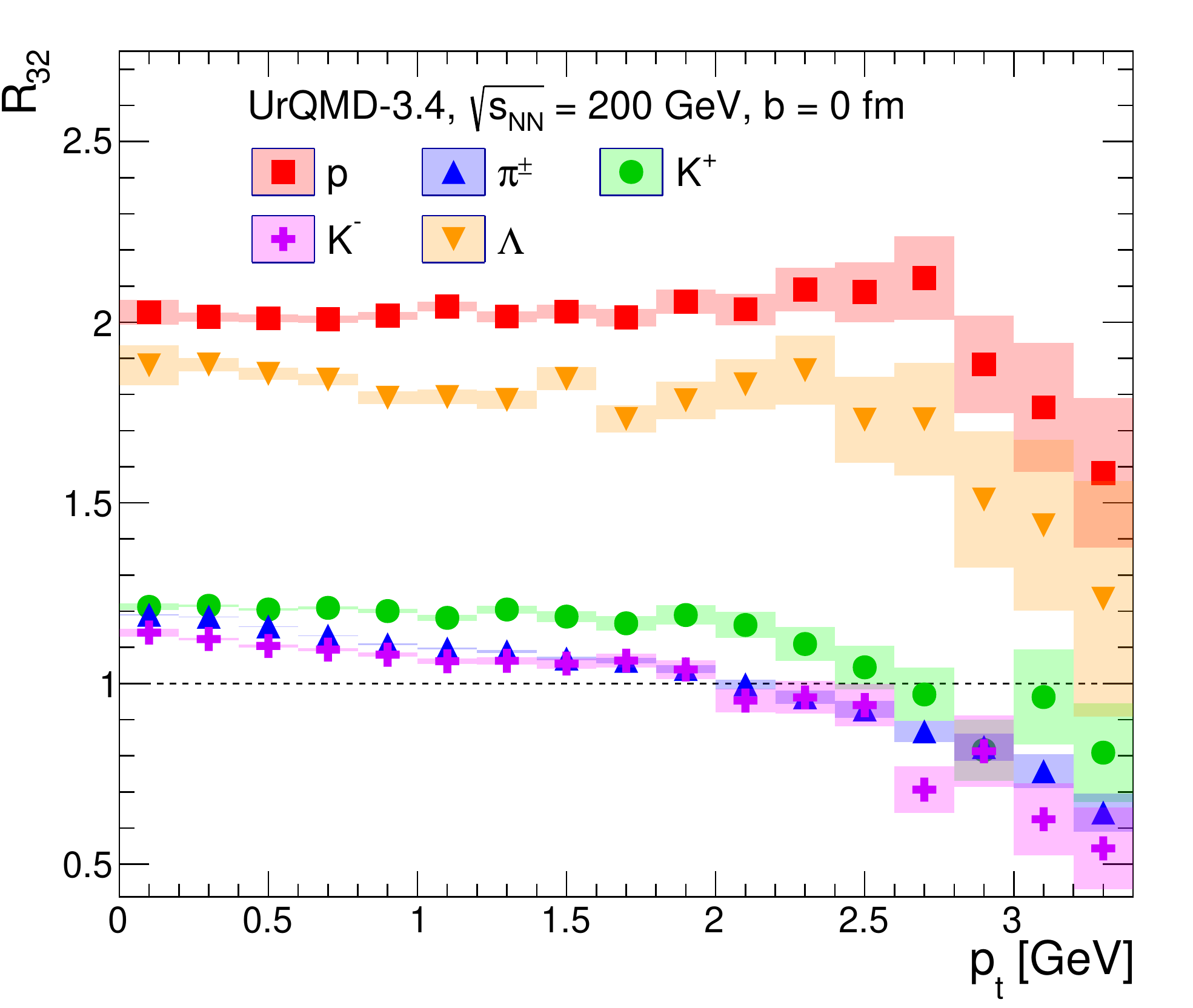}}
        \caption{\small {\bf Left panel:} The ratio of hadronic yields per rapidity unit $\frac{d N}{d y}$ expected for the most central and simultaneous Pb+Pb+Pb TNC to the one for the most central Pb+Pb collisions (i.e. 3-to-2 nuclei enhancement factor for yields) for the collision energy $\sqrt{s_{NN}} = 200$ GeV. {\bf Right panel:} Ratio of transversal momentum spectra of the most central and simultaneous Pb+Pb+Pb TNC to the one found for the most central Pb+Pb collisions (3-to-2 nuclei enhancement factor for $p_T$ spectra) of hadrons obtained for the same collision energy $\sqrt{s_{NN}} = 200$ GeV as a function of particle transverse momentum found at $\left|y\right| < 0.1$.
        }
        \label{fig1_Bugaev} 
\end{figure}

After about three decades of investigating the phase diagram of strongly interacting matter in binary nuclear (A+A) collisions
it became clear that the most interesting phenomena such as the expected chiral symmetry restoration and the deconfinement phase transitions may occur at rather low center-of-mass collision energies with the thresholds $\sqrt{s_{NN}} \simeq 4-5$ GeV \cite{Bugaev_Ref1,Bugaev_Ref2,Bugaev_Ref3,Bugaev_Ref4,Bugaev_Ref5} for the chiral symmetry restoration phase transition
and $\sqrt{s_{NN}} \simeq 9-10$ GeV \cite{Bugaev_Ref1,Bugaev_Ref2,Bugaev_Ref3,Bugaev_Ref4,Bugaev_Ref5} for the deconfinement one. Perhaps, the hardest lesson that the community of heavy-ion collisions learned after so many years is that in addition to the A+A collisions we need an independent and reliable source of information about the equation of state (EoS) of strongly interacting matter \cite{Bugaev_Ref6,Bugaev_Ref7}. Sharing this idea, we suggest to consider  the triple nuclear collisions (TNC) \cite{Bugaev_Ref8} as an additional and independent source of information. The TNC can be done either by inserting a super-thin target into the interaction zone of two colliding beams, or by making the jet target consisting of small metallic droplets of about 1-2 $\mu$m size, or by installing the third ring in a perpendicular direction to two colliding beams \cite{Bugaev_Ref8,Bugaev_Ref9}. Since the TNC is a very fresh topic, in this short work we just outline its principal advantages over the binary nuclear collisions (BNC), whereas the estimates of TNC rates and details of the experimental setup we will discuss in the separate publication \cite{Bugaev_Ref9}.


\section{New Extreme Conditions achieved in TNC}
\label{sect2_Bugaev}

In order to model the TNC,  we used the UrQMD 3.4 model \cite{Bugaev_Ref10,Bugaev_Ref11}, which is able to rather satisfactory describe the bulk data on secondary hadron yields and their transversal momentum spectra. The primary goal of this work is to demonstrate the entirely new possibilities to study the QCD phase diagram with TNC at  essentially higher baryonic and electric charge densities than the ones achieved in A+A collisions. Therefore, here we concentrate on the bulk properties of baryon production, while the question of possible phase transformations at low energies of collisions we leave for further exploration by more specialized approaches.

\begin{figure}[t]
        \centerline{
                \includegraphics[width=60mm,height=56mm]{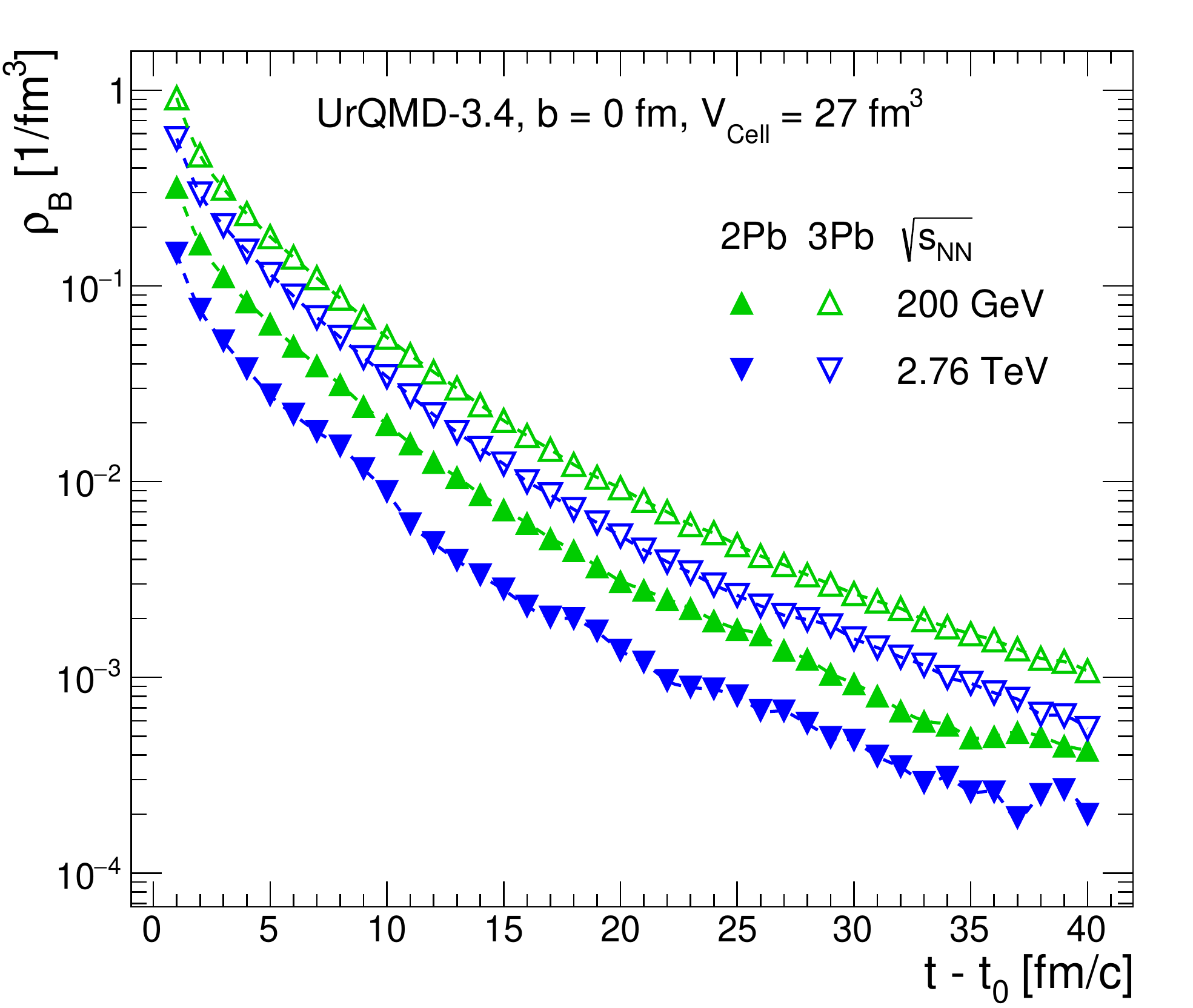} \hspace*{1.1mm}
                \includegraphics[width=70mm,height=56mm]{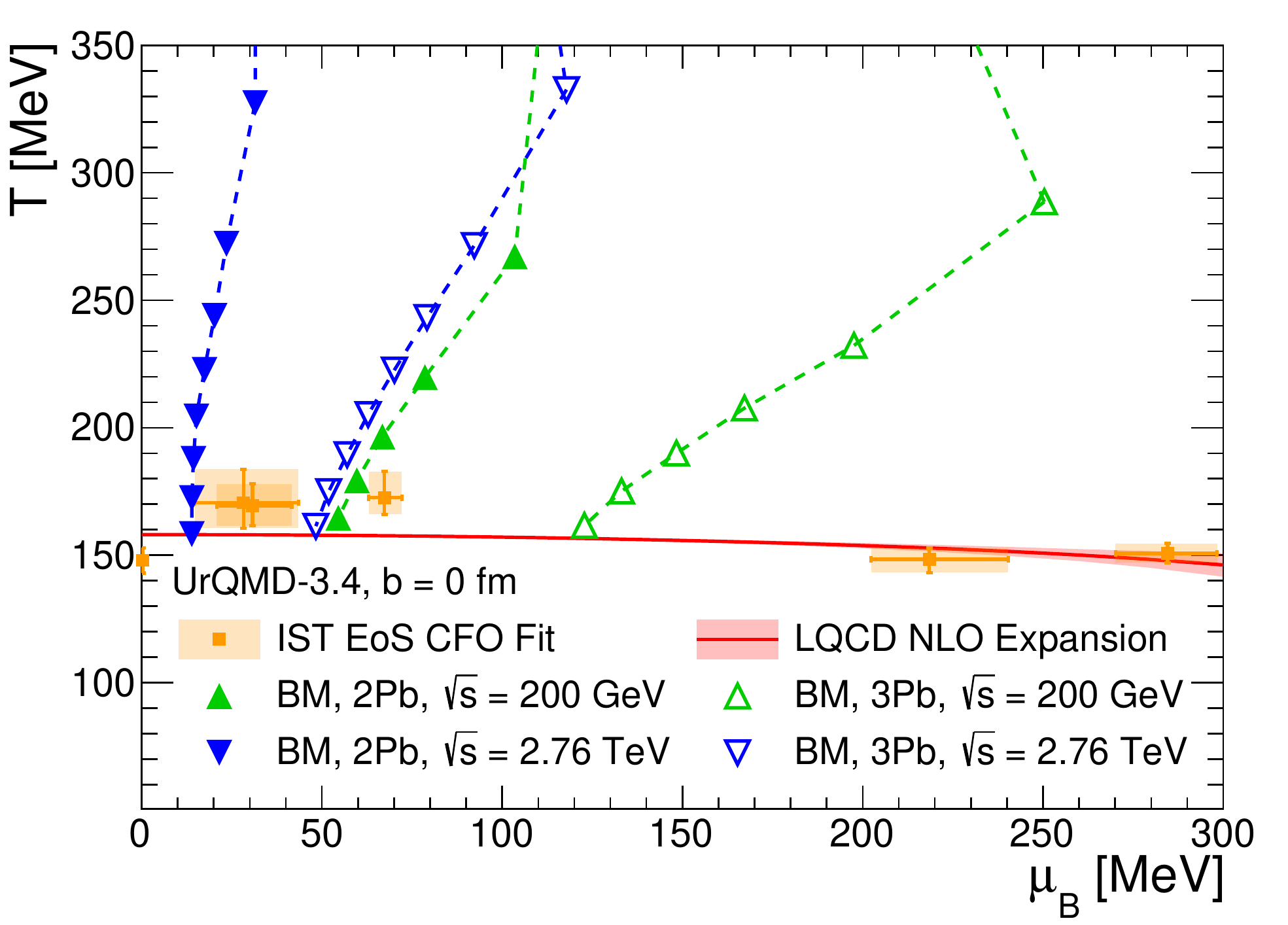}}
        \caption{\small {\bf Left panel:} The time evolution of the baryonic charge density in the central cell during the process of ordinary BNC (filled symbols) and for the TNC (empty symbols) found for $\sqrt{s_{NN}} =200$ GeV (squares) and for $\sqrt{s_{NN}} =2.76$ TeV (circles). The time $t_0$ is the moment, when the remnants of projectile nuclei have passed through the central cell.
        {\bf Right panel:} The evolution of central cell parameters in the $\mu_B-T$ plane obtained with the MIT Bag Model equation of state \cite{BagModelSM}. The filled symbols correspond to the Pb+Pb collisions, while the empty ones to the Pb+Pb+Pb TNC. Collision energy $\sqrt{s_{NN}} = 200$ GeV ($\sqrt{s_{NN}} =2.76$ TeV) points are shown by the triangles up (down). The topmost points correspond to the time $t-t_0 > 1$ fm. The curve of pseudo-critical temperature corresponds to a lattice QCD parameterization \cite{lQCDparam}, while the crosses correspond to the chemical freeze-out parameters in Pb+Pb collisions found in Ref. \cite{IST1}.
        }
        \label{fig2_Bugaev} 
\end{figure}

In order to demonstrate the strength of the effect, in Fig.~\ref{fig1_Bugaev} we show the 3-to-2 enhancement factor as a ratio of the quantity obtained by UrQMD 3.4 model for the most central Pb+Pb+Pb to the same quantity found for the most central Pb+Pb collisions at the same energy of colliding beams $\sqrt{s_{NN}} = 200$ GeV. As one can see from the left panel of Fig.~\ref{fig1_Bugaev}, the yield of protons and $\Lambda$-hyperons in the TNC is enhanced almost by a factor of 2 and, hence, this feature can be, in principle, used to detect the TNC in the event-by-event analysis. Another distinct difference between TNC and BNC can be seen in the right panel of Fig.~\ref{fig1_Bugaev}. The number of pions and kaons is slightly enhanced at $p_T \le 2$ GeV, while their number is suppressed for $p_T > 2.5$ GeV. A qualitatively similar enhancement for $p_T \le 3.5$ GeV and suppression for $p_T > 4$ GeV we found for protons and $\Lambda$-hyperons in the TNC with the difference, that both the enhancement and the suppression of $p_T$ spectra of these particles are much stronger than for pions and kaons. Therefore, the ratio of $\frac{d^2 N}{d y d p_T }$ for $p_T \le 2$ GeV to the same quantity, but for $p_T > 5$ GeV measured in the TNC and in the BNC will be much larger for the TNC.

This effect of the transverse momentum redistribution in TNC we observed even in the p+C+p collisions \cite{Bugaev_Ref9}, when the proton-proton collisions occur inside the carbon nucleus. Because of the presence of surrounding nucleons the products of the primary p+p or p+n collisions that are moving in the transversal direction re-scatter on the nucleons of the target nucleus and lose their transverse momenta and, moreover, they produce an additional amount of secondary hadrons in comparison with the ordinary BNC. Therefore, the probability to detect the slow-moving nucleons at the midrapidity increases essentially. On the other hand, after one or two re-scatterings the products of primary collisions with the large $p_T$ value lose the part of their momentum and increase the number of states with low $p_T$ values. As one can see from the right panel of Fig.~\ref{fig1_Bugaev}, in the TNC the effect of transverse momentum redistribution of protons and $\Lambda$-hyperons is rather strong and, hence, it can be used to identify the Pb+Pb+Pb collisions and even the p+C+p ones in the experiments.

Our simulations with UrQMD 3.4 model show us that using the TNC one can probe the new extreme conditions, namely very high baryonic and electric charge densities, even at highest RHIC energies of collisions. To demonstrate this we studied the evolution of the central cell with the size $3\times 3\times 3 $ fm$^3$. The results for the most central Pb+Pb+Pb TNC for $\sqrt {s_{NN}}$ = 200 GeV (RHIC) and $\sqrt {s_{NN}}$ = 2.76 TeV (LHC) are shown in Fig.~\ref{fig2_Bugaev}. From the left panel of Fig.~\ref{fig2_Bugaev} it is easy to see that compared to the usual BNC the TNC provide an essential increase of the initial baryonic charge density: from 0.321 fm$^{-3}$ to 0.913 fm$^{-3}$ for $\sqrt {s_{NN}}$ = 200 GeV and from 0.146 fm$^{-3}$ to 0.567 fm$^{-3}$ for $\sqrt {s_{NN}}$ = 2.76 TeV. From these results it is clear that in the TNC one can expect a formation of about 3 times higher initial baryonic charge density for the RHIC top energy and 4 times denser ones for the LHC energy than in the BNC. The time $t_0 \simeq 10$ fm/c for the Pb+Pb+Pb TNC and $t_0 \simeq 1$ fm/c for Pb + Pb collisions correspond to the time when the remnants of target nuclei have passed through the central cell. To quantify the values of baryonic chemical potential $\mu_B$ and temperature $T$ inside the central cell, we used the MIT Bag Model equation of state \cite{BagModelSM} for 3 colors and 3 massless quark flavors, i.e. for the pressure $p^{BM} = \frac{95}{180}\pi^2 T^4 + \frac{T^2 \mu_B^2}{6} + \frac{\mu_B^4}{108 \, \pi^2} - B_{vac} $, with a conservative value $B_{vac}^\frac{1}{4} = 206$ MeV \cite{WongBook}. Equating the baryonic charge density $\rho_{cell}$ and the energy density $\epsilon_{cell}$ found for the central cell to the corresponding quantities of the MIT Bag Model $\rho_{BM} \equiv \frac{\partial p^{BM} }{\partial \mu_B}$ and $\epsilon_{BM} = T\frac{\partial p^{BM} }{\partial T} +\mu_B \rho_{BM} - p^{BM}$, we found the values of $\mu_B$ and $T$ in central cell (see Fig.~\ref{fig2_Bugaev}). As one can see from the right panel of Fig.~\ref{fig2_Bugaev}, for the same temperature the values of $\mu_B$ are essentially larger in the TNC than in the BNC.

\section{Conclusions}
\label{Conclusions_Bugaev}
In this work,  we briefly showed that for the same energy of colliding beams, in TNC the QCD matter can reach about 3 times higher baryonic charge density than in BNC. Apparently, this opens entirely new perspectives to study the QCD matter phase diagram under new extreme conditions and, hence, we believe that after developing the TNC concept, future experiments will be done with the TNC.

\begin{acknowledgement}
        O.V.V., V.M.P. and K.A.B. acknowledge a partial support from the Program “Participation in the novel international projects in high energy and nuclear physics" launched by the Section of Nuclear Physics and Energetics of NAS of Ukraine. The work of L.V.B. and E.E.Z. was supported by the Norwegian Research Council (NFR) grant No. 255253/F50 - “CERN Heavy Ion Theory" and by the Russian Foundation for Basic Research under the grants No. 18-02-40084 and 18-02-40085. A.T received funding from the Russian Foundation for Basic Research (RFBR) under the project No. 18-02-40086. The simulations were done on the JINR supercomputer Govorun.

\end{acknowledgement}

%

\begin{thebibliography}{}
        %
        %

        \bibitem{Bugaev_Ref1}
        %
        K. A. Bugaev et al.,
        Phys. Part. Nucl. Lett. {\bf 12}, 238 (2015).

        \bibitem{Bugaev_Ref2}
        %
        K. A. Bugaev et al.,
        Eur. Phys. J. A {\bf 52} (6), 175 (2016) .

        \bibitem{Bugaev_Ref3}
        %
        W. Cassing, A. Palmese, P. Moreau, and E. L. Bratkovskaya,
        Phys. Rev. C {\bf 93}, 014902 (2016).

        \bibitem{Bugaev_Ref4}
        %
        A. Palmese et al.,
        Phys. Rev. C {\bf 94}, 044912 (2016).

        \bibitem{Bugaev_Ref5}
        K.~A.~Bugaev et al.,
        Phys. Part. Nucl. Lett. \textbf{15}, 210 (2018).


        \bibitem{Bugaev_Ref6}
        %
        E. R. Most et al., Phys. Rev. Lett. {\bf 122} , 061101 (2019).

        \bibitem{Bugaev_Ref7}
        %
        A. Bauswein et al., Phys. Rev. Lett. {\bf 122}, 061102 (2019).

        \bibitem{Bugaev_Ref8}
        %
        K. Bugaev et al., talk at the Online «Strangeness in Quark Matter» Conference 2021, Brookhaven, May 17-22, 2021; https://indico.cern.ch/event/985652/sessions/392917/\#20210521

        \bibitem{Bugaev_Ref9}
        %
        K. Bugaev et al., talk at the Offshell-2021 — The virtual HEP conference on Run4 at LHC, July 6-9, 2021;
        https://indico.cern.ch/event/968055/contributions/4273917/

\bibitem{Bugaev_Ref10}
%
S.A. Bass et al., Prog. Part. Nucl. Phys.  {\bf 41},   255 (1998). 

\bibitem{Bugaev_Ref11}
%
 M. Bleicher et al., J. Phys. G {\bf 25}, 1859 (1999).
 
 \bibitem{BagModelSM}  
%
A. Chodos, R. L. Jaffe, K. Johnson, C. B. Thorn and  V. F. Weisskopf,
Phys. Rev. D {\bf 9}, 3471   (1974).

\bibitem{lQCDparam}  
%
 S. Borsanyi {\it et al.}, Phys. Rev. Lett. {\bf 125}, 052001  (2020).
 
 \bibitem{IST1}  
%
V. V. Sagun {\it et al.},
 Eur. Phys. J. A {\bf 54}  (6), 100 (2018). 
 
 \bibitem{WongBook}  
%
C. Y. Wong, {\it "Introduction to high-energy heavy ion collisions,"} (Singapore,
World Scientific, 1994)

\end{thebibliography}
%
%

\end{document}